%% file: glimpse_paper1.tex
\documentclass[10pt,preprint]{aastex}

\newcommand{\glimpse} {\mbox{\it GLIMPSE~}}
\newcommand{\sirtf}{\mbox {\it SIRTF~}}
\newcommand{\irac}{\mbox {\it IRAC~}}
\renewcommand{\deg}{$^\circ$}

\def\etal{\hbox{\it et al.}}

\slugcomment{Accepted for publication in PASP}

\shorttitle{\glimpse: A \sirtf Legacy Project}
\shortauthors{Benjamin et al. }

\begin{document}

%% LaTeX will automatically break titles if they run longer than
%% one line. However, you may use \\ to force a line break if
%% you desire.

\newlength{\hugepicsize}
\setlength{\hugepicsize}{5.5in}
\newlength{\bigpicsize}
\setlength{\bigpicsize}{4.5in}
\newlength{\smpicsize}
\setlength{\smpicsize}{3.5in}
\newlength{\tinypicsize}
\setlength{\tinypicsize}{2.5in}
\newlength{\tinierpicsize}
\setlength{\tinierpicsize}{2.0in}

\title{\glimpse: I.  A \sirtf Legacy Project to Map the Inner Galaxy}

%% Use \author, \affil, and the \and command to format
%% author and affiliation information.
%% Note that \email has replaced the old \authoremail command
%% from AASTeX v4.0. You can use \email to mark an email address
%% anywhere in the paper, not just in the front matter.
%% As in the title, you can use \\ to force line breaks.

\author{Robert A. Benjamin\altaffilmark{1}, 
E. Churchwell\altaffilmark{2},
Brian L. Babler\altaffilmark{2}, 
T. M. Bania\altaffilmark{3},    
Dan P. Clemens\altaffilmark{3}, 
Martin Cohen\altaffilmark{4}, 
John M. Dickey\altaffilmark{5},  
R\'emy Indebetouw\altaffilmark{2}, 
James M. Jackson\altaffilmark{3}, 
Henry A. Kobulnicky\altaffilmark{6}, 
Alex Lazarian\altaffilmark{2}, 
A. P. Marston\altaffilmark{7}, 
John S. Mathis\altaffilmark{2}, 
Marilyn R. Meade\altaffilmark{2}, 
Sara Seager\altaffilmark{8}, 
S. R. Stolovy\altaffilmark{7}, 
C. Watson\altaffilmark{2},   
Barbara A. Whitney\altaffilmark{9}, 
Michael J. Wolff\altaffilmark{9},   
and 
Mark G. Wolfire\altaffilmark{10}}

\altaffiltext{1}{University of Wisconsin-Madison, Dept. of Physics, 1150 University Ave., Madison, WI 53706; benjamin@physics.wisc.edu}
\altaffiltext{2}{University of Wisconsin-Madison, Dept. of Astronomy, 475 N. Charter St., Madison, WI 53706}
\altaffiltext{3}{Institute for Astrophysical Research, Boston University, 725 Commonwealth Avenue, Boston, MA 02215}
\altaffiltext{4}{Radio Astronomy Laboratory, University of California, Berkeley, CA 94720}
\altaffiltext{5}{Astronomy Department, University of Minnesota, 116 Church Street SE, Minneapolis, MN 55455}
\altaffiltext{6}{Department of Physics and Astronomy, University of Wyoming, Laramie, WY 82071}
\altaffiltext{7}{SIRTF Science Center, California Institute of Technology, MS 220-6, Pasadena, CA 91125}
\altaffiltext{8}{Carnegie Institution of Washington, Department of Terrestrial Magnetism, 5241 Broad Branch Rd NW, Washington, D.C. 20015}
\altaffiltext{9}{Space Science Institute, 3100 Marine Street, Suite A353, Boulder, CO 80303-1058}
\altaffiltext{10}{Department of Astronomy, University of Maryland, College Park, MD 20742}

\begin{abstract}

{\it GLIMPSE} (Galactic Legacy Infrared Mid-Plane Survey
Extraordinaire), a \sirtf Legacy Science Program, will be a fully
sampled, confusion-limited infrared survey of
the inner two-thirds of the Galactic disk with a pixel resolution of
$\sim 1.2^{\prime \prime}$ using the Infrared Array Camera (\irac) at 3.6, 4.5, 5.8, and 8.0 ${\rm \mu
m}$. The survey will cover Galactic latitudes $|b| \le 1^{\circ}$ and longitudes
$|l|=10^{\circ}$ to $65^{\circ}$ (both sides of the Galactic center). The
survey area contains the outer ends of the Galactic bar, the Galactic
molecular ring, and the inner spiral arms. The \glimpse team will
process these data to produce a point source catalog, a point source data archive, and a set of
mosaicked images. We summarize our observing strategy, give details of
our data products, and summarize some of the principal science
questions that will be addressed using \glimpse data. Up-to-date documentation,
survey progress, and information on complementary datasets are
available on the \glimpse web site: www.astro.wisc.edu/glimpse.
\end{abstract}

%% Keywords should appear after the \end{abstract} command. The uncommented
%% example has been keyed in ApJ style. See the instructions to authors
%% for the journal to which you are submitting your paper to determine
%% what keyword punctuation is appropriate.

\keywords{infrared:general -- infrared:ISM -- infrared:stars -- survey -- stars:general -- ISM:general -- Galaxy:structure -- Galaxy:stellar content }

%% From the front matter, we move on to the body of the paper.
%% In the first two sections, notice the use of the natbib \citep
%% and \citet commands to identify citations.  The citations are
%% tied to the reference list via symbolic KEYs. The KEY corresponds
%% to the KEY in the \bibitem in the reference list below. We have
%% chosen the first three characters of the first author's name plus
%% the last two numeral of the year of publication as our KEY for
%% each reference.
\section{Motivation for \glimpse}

The inner workings of our own Galaxy are as mysterious as those
of galaxies located millions of light years away, mainly because of
our unfavorable location in the mid-plane outskirts of the Milky Way's
dusty disk. The structure of the Galactic disk has been determined
primarily from the distributions of atomic hydrogen (Westerhout 1957) and carbon
monoxide   which together contain no more than about 10\% of the visible
mass of the Galaxy (Scoville \& Solomon 1975).  The Galaxy is a typical
luminous spiral, but its stellar distribution, particularly in
the inner quadrants, is poorly known.  For example, although there is
significant evidence that the Galaxy has a molecular ring (Scoville \& Solomon 1975), the
number of stars recently formed in this ring, and the resultant appearance of
the Galaxy's spiral arms to an outside observer, are unknown.  

The principal impediment to cataloging the stellar content of the
inner Galaxy has been dust obscuration of the visible light from
stars. What has been needed is a survey with high
sensitivity and angular resolution in the middle infrared and longer wavelengths. The recently
completed 2MASS (Two Micron All-Sky Survey) survey (Cutri et al. 2001)
has been an important step in this process, producing a view of the
inner Galaxy at wavelengths as long as 2.2 ${\rm \mu m}$. Yet even at
these wavelengths, the extinction due to dust significantly
compromises our ability to probe the stellar content of the inner
Galaxy and obtain accurate measurements of fundamental Galactic
parameters.

\glimpse (Galactic Legacy Infrared Mid-Plane Survey Extraordinaire), a
Legacy Project using {\sirtf} (Space InfraRed Telescope Facility, see
Gallagher, Irace, \& Werner 2002 for a description of the full
facility), is a project to map the infrared emission from the inner
Galaxy over the two strips $|b| \le 1^{\circ}$ and $|l|=10^{\circ}$ to
$65^{\circ}$ using the IRAC instrument (see Fazio et al. 1998).  Of all the \sirtf Legacy programs, this survey will
cover the largest area on the sky (some 220 square degrees) and will yield the most panoramic
images.  With a total observing time of approximately
400 hours, the survey will consist of over 80,000 pointings, each
resulting in four simultaneous IRAC  images at 3.6
${\rm \mu m}$, 4.5 ${\rm \mu m}$, 5.8 ${\rm \mu m}$, and 8.0 ${\rm \mu
m}$. The \glimpse team will use these data to produce a highly
reliable point source catalog, a somewhat deeper point source archive, a set of
mosaicked images, and associated analysis software.

The survey area was chosen to include all of the major known or
suspected stellar components of the inner Galaxy (except the central
bulge), namely the outer ends of the Galactic bar, the molecular ring at a Galactocentric radius of  $\sim 3-5$ kpc, the inner disk, and the inner spiral
arms and spiral arm tangencies (See Figure 1). The inner $ \pm
10^{\circ}$ degrees of the Galaxy are excluded from our survey because
of the high background and confusion present there. 

The \glimpse team will focus on two central science questions:

\begin{itemize}{}

\item \underline{{\it What is the structure of the inner Galaxy?}} What is the structure of the disk and 
molecular ring? What are the number and locations of spiral arms? What is the nature of the central bar as
traced by the spatial distribution of stars and infrared-bright star
formation regions? In particular, we will address the question of
whether the Galaxy is a ringed galaxy by correlating the stellar
content with studies of the molecular ring of the Galaxy (Clemens et
al. 2000).

\item \underline{{\it What are the statistics and physics of star formation?}} How does the nature of star formation
depend on mass, stage of evolution, and location in the Milky Way? What will an unbiased infrared  survey with well over 2000 star formation regions reveal about the earliest evolutionary stages of star formation? How does the infrared emission change during each of the principal stages of star formation?  
\end{itemize}

%\begin{figure}
%\figurenum{1}
%\plotone{fg1_paper1_lowres.eps}
\begin{figure}[ht!]
\begin{center}
\includegraphics[totalheight=\smpicsize]{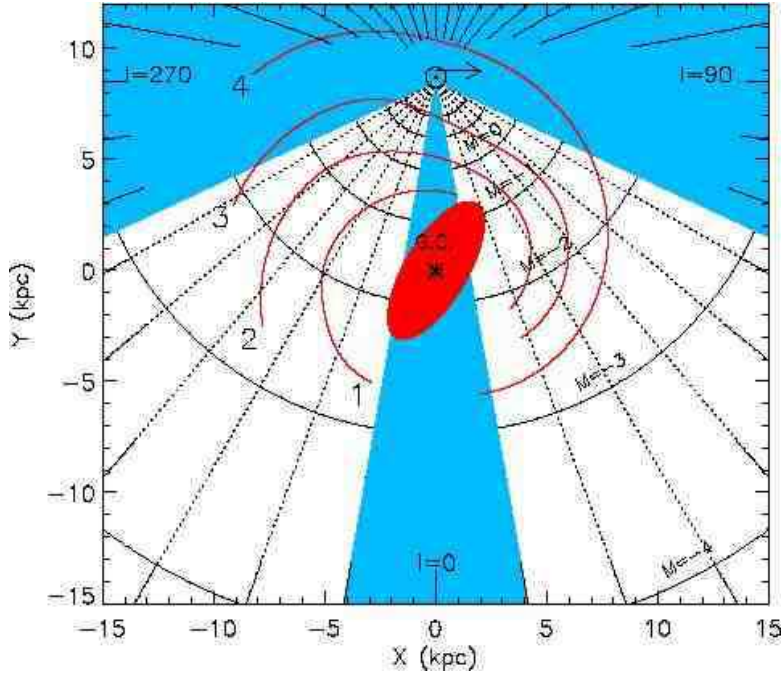}
\caption{\small{The Galactic Plane, with the Sun at the top, and the direction of solar motion noted with an arrow. The area
covered by \glimpse is unshaded.  The dotted lines are intervals of
10\deg\ in longitude. The solid circles indicate the 5$\sigma$
detection distance for objects with absolute magnitude $M_{8\mu
m}=2,1,..-4$ assuming no extinction. Young OB stars ($M=-4.35$) and M
giants ($M=-4.24$) will be visible throughout the Galaxy.  The approximate
positions of Galactic spiral arms (Taylor \& Cordes 1993) are
indicated with bold lines and are numbered: (1) Norma Arm, (2)
Scutum-Crux Arm, (3) Sagittarius Arm, and (4) Perseus Arm. The central
oval represents the approximate extent of the central bar (Gerhard
2002; Cole \& Weinberg 2002) with the Galactic Center marked with an
asterix.}}
\end{center}
\end{figure}

The major purpose for \glimpse is to provide the community with a Legacy dataset complete to a well-defined flux limit suitable for a wide variety of astrophysical investigations. These might encompass population studies of
different classes of Galactic objects, including regions of low and
high mass star formation, highly evolved AGB and OH/IR stars, stars
with circumstellar dust shells, cool stars of all luminosities,
photodissociation regions (PDRs), proto-planetary nebulae, planetary
nebulae, Wolf-Rayet stars, open clusters, and supernova
remnants. Table 1 lists several classes of objects and the number of
these objects already known to exist in the \glimpse survey area. \glimpse data will also be used to study the infrared dark clouds (Egan et al. 1998) revealed by MSX (Midcourse Space Experiment) and ISO (Infrared Space Observatory) and will extend the catalog of these objects to smaller sizes and fainter limits.

We expect that \glimpse data will be used by the community for investigations that we
can not anticipate. Of all the directions in the sky, the inner Galaxy
has heretofore been the most inaccessible because of dust
obscuration. The \glimpse program will reveal for the first time a wealth of
completely new stars, clusters, and galaxies.  It is this element of
serendipity that makes \glimpse a particularly exciting endeavor!

In this paper, we provide a description of the  \glimpse observing plan (\S 2.1), data processing (\S 2.2), and data products (\S 2.3). Next, we give details
on the scientific goals and challenges of the \glimpse survey,
including high mass star formation in the inner Galaxy (\S 3.1.1),
Galactic structure (\S 3.1.2), Legacy science (\S 3.2), and a  description of the data sets that complement \glimpse
(\S 3.3). A summary is given in \S 4. 

\section{Project Description}

\sirtf and the  Infrared Array Camera (IRAC) will be used to image  two long strips comprising 220 square
degrees at wavelengths centered on 3.6, 4.5, 5.8, and 8.0 $\mu$m. The area surveyed by \glimpse ($|b| \le 1^{\circ}$,$|l|=10^{\circ}-65^{\circ}$) contains most of the star formation activity of the Galaxy, the outer ends of the central bar, all of the Galactic molecular ring, and four spiral arm tangencies. The principal characteristics of \glimpse are listed in
Table 2. The improvements in sensitivity, angular resolution, and areal coverage
afforded by \glimpse over previous infrared surveys of the Galactic plane are shown in Figure
2.

%\begin{figure}
%\figurenum{2a}
%\plotone{fg2a_paper1_lowres.eps}
\begin{figure}[ht!]
\begin{center}
\includegraphics[angle=0,totalheight=\tinierpicsize]{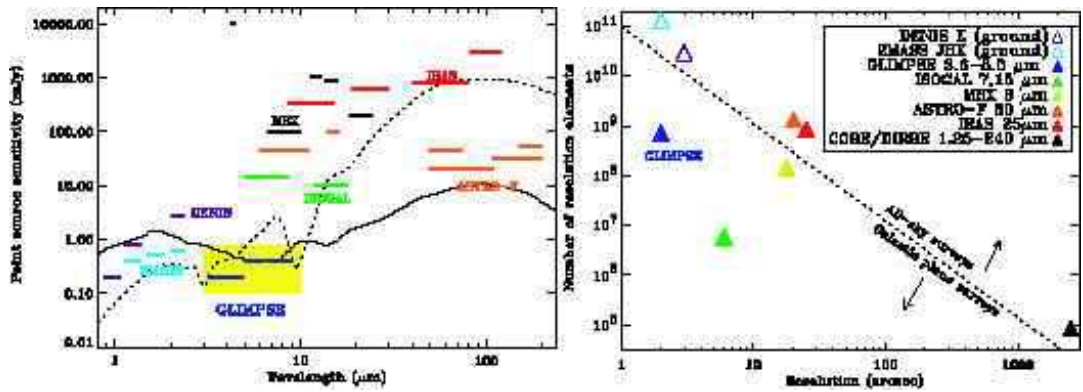}
\caption{\small{A comparison of the \glimpse sensitivity limit to the sensitivity of other ground and space-based infrared surveys. This shows the good match in sensitivity between \glimpse and 2MASS. The curves show model spectra of Whitney et al (2003) for a 1 $L_{\sun}$ T Tauri star at a distance of 0.7 kpc (solid) and a deeply embedded 1 $L_{\sun}$ protostar at a distance of 0.6 kpc(dotted). The number of resolution elements versus resolution for \glimpse and other infrared surveys. \glimpse will have the largest number of resolution elements of any Galactic plane-only survey, and a comparable number of resolution elements to several space-based all-sky infrared surveys.
}}
\end{center}
\end{figure}

The \glimpse team will provide the following products: a high reliability
\glimpse Point Source Catalog (GPSC) containing about 10 million objects, a \glimpse Point Source Archive
(GPSA; $5\sigma$), and a Mosaicked Image Atlas of the entire surveyed
area in all four IRAC bands. All these data products will be made available via the 
SSC (\sirtf Science Center). In addition, a set of web-accessed modeling tools will
permit users to  interpret \sirtf and
other IR data. 

Here we discuss the \glimpse implementation, data processing plans, and
 resulting data products. Up-to-date information on the progress of \glimpse observations (including a graphical survey tracker), data reduction and
data releases can be found at the \glimpse website:
www.astro.wisc.edu/glimpse.

\subsection{\glimpse implementation}

The orbit, orientation, and viewing limits on \sirtf conspire to make mapping the Galactic plane a complex process. Furthermore, the spectacular sensitivity of IRAC/SIRTF means that only exceedingly short exposures of the Galactic plane will not be saturated. These issues drive much of the \glimpse observing implementation. 

{\it Survey strategy:} During standard \sirtf operations, IRAC ``campaigns,'' lasting 10 days to two weeks,
will be scheduled. A fraction of these IRAC campaigns during the first year will be devoted to the \glimpse program which will observe many IRAC frames tiled together into  
$\sim 15^{\circ} \times 2^{\circ}$  segments
 of the Galactic plane. Each 
 $\sim 30$ square degree segment will
consist of 35-45  ``chained'' AORs (Astronomical Observing
Requests).\footnote{``Chained'' AORs must be carried out within a certain
amount of time of each other. For \glimpse, the typical maximum
separation between AORs is 3 hours.} Each AOR will cover a narrow
rectangular strip $0.3^{\circ} \times (2-3)^{\circ}$ spanning $b\cong -1^{\circ}$ to $b\cong+1^{\circ}$, but inclined to the Galactic plane.  The inclination angle of
the AOR to the Galactic plane depends upon the spacecraft roll angle. For \glimpse
observations, the roll angle will lie between 15-50$^{\circ}$ from Galactic North.  Example AOR sky coverage in the direction
of the star formation region W51 is shown in Figure 3.

%\begin{figure}
%\figurenum{3}
%\plotone{fg3.eps}
\begin{figure}[ht!]
\begin{center}
\includegraphics[angle=0,totalheight=\bigpicsize]{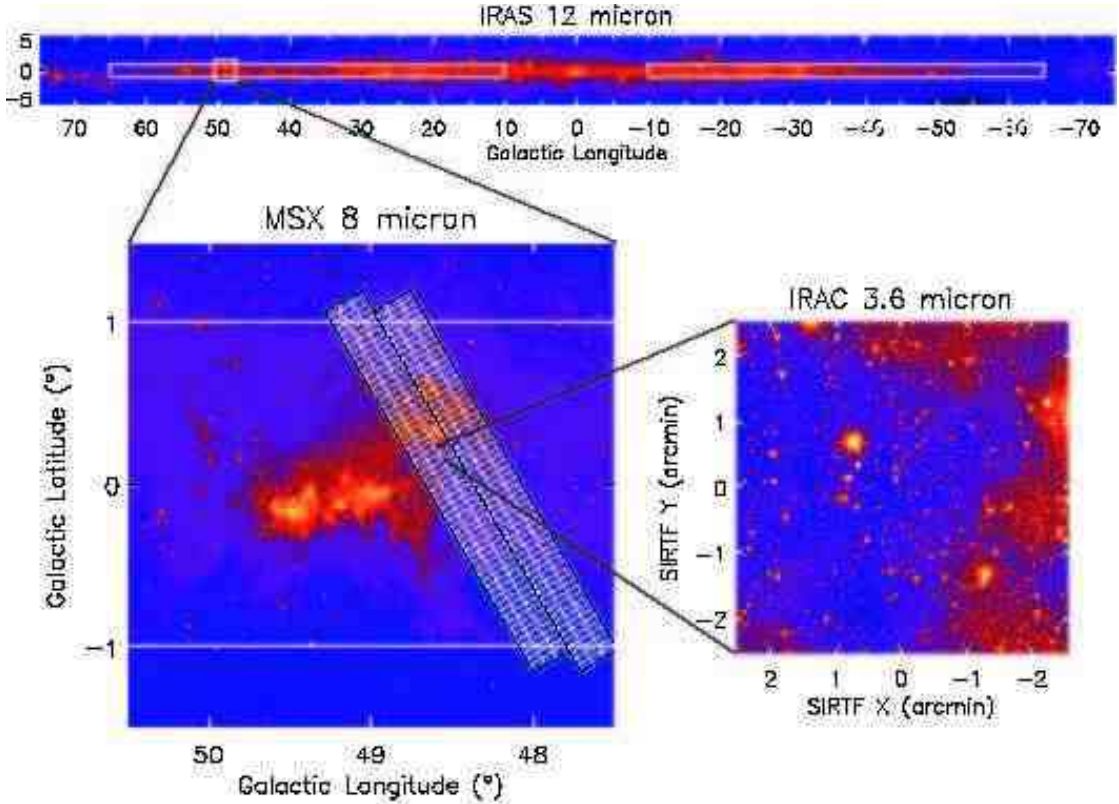}
\caption{\small{The upper figure shows an IRAS 12 $\mu$m image of the Galactic plane  with long rectangular boxes indicating the region to be covered by \glimpse. The subregion on the lower left is an MSX $8 \mu$m map (Price et al. 2001) of the star formation region W51. This map shows the location of a set of tiled IRAC frames across the image. The black rectangle represents one Astronomical Observing Request (AOR) consisting of about 180 individual $5\arcmin \times 5\arcmin$ frames. A 15$^{\circ}$ segment in longitude will typically require about 40 AORs. The subregion of this plot is a synthetic \glimpse  IRAC 3.6$\mu$m frame (1.2\arcsec pixel resolution), containing all the 2MASS point sources in this direction, supplemented with additional, fainter sources using the SKY model described in Wainscoat et al. (1992). The diffuse flux is taken from MSX observations.}}
\end{center}
\end{figure}

Each IRAC pointing simultaneously images  two adjacent $5.17\arcmin \times 5.17\arcmin$ fields in two bands. An IRAC frame has 256 $\times$ 256 pixels; the 3.6 and 5.8 $\mu$m fields coincide on the sky and the 4.5 and 8.0 $\mu$m fields coincide on the sky, but the frame edges of the two fields are separated by 1.5$\arcmin$.  The total integration time per position is two seconds.  The
observations will be stepped by half-frames (128 pixels; 2.58$\arcmin$) along the long axis of the
AOR (which could be the spacecraft X or Y-axis, depending upon the roll
angle). Every sky direction in the \glimpse region will be visited at least twice.  The time separation between the two visits will range from 20
seconds (the time between pointings) to 3 hours (the time between AORs). The
frame overlap along the short axis of the AOR will be $14.4\arcsec$
(12 pixels). A typical AOR will take 1.5 hours to complete, yielding
$(3-4) \times (47-70)$ IRAC frames per band. Each $\sim 15^{\circ}
\times 2^{\circ}$ segment of the Galactic plane will require 35-45
AORs; the entire survey will require 8 segments.  

{\it Observing Strategy Validation:} Time will be scheduled during early \sirtf operations to validate our observing strategy (OSV). The \glimpse OSV region will be chosen to sample a range of 
 stellar densities and diffuse
background levels characteristic of the inner Galactic plane. Regions of particularly high stellar density and diffuse background will be included to assess our strategy in the most challenging cases. The OSV will view either W51,  G333.3-0.4, G305.2+0.2, or NGC6334, depending upon \sirtf launch date.  The goal of these observations
is to assure that \glimpse data will have the highest reliability
point-source extraction and the greatest possible sky-coverage, and to
prevent the occurrence of gaps in the survey area. \glimpse reliability and completeness calculations, which have been based on simulated data, will be tested by these observations.  The OSV
will consist of four $1^{\circ} \times 0.17^{\circ}$ AORs nearly identical to normal \glimpse AORs. These will be repeated with varying frame offsets (along the strip direction) and overlaps (between strips). Finally, \glimpse OSV will perform a single $2^{\circ}
\times 0.17^{\circ}$ AOR to allow us to evaluate the pointing reconstruction accuracy for the full half hour
that it takes to do a single diagonal band across the Galactic plane, testing roll angle uncertainties between AORs, and determining criteria for establishing reliability and completeness.  At the end of the OSV period, the outer longitude limits of the survey may be adjusted so that the total survey time does not exceed 400 hours. 

{\it Frame Details:} The simulated 3.6$\mu$m IRAC frame for the W51
region shown in Figure 3 illustrates both the wealth of detail that
\glimpse will uncover and the significant challenges for processing and
analyzing these data.  In this region,
there are 400-600 2MASS sources per frame, all of which are detectable
by \glimpse. (See Table 3 for the 5$\sigma$ sensitivities and
estimated flux limits for 99.5\% reliability.) Examination of 2MASS and MSX data for this region indicates that we should expect 2-4 saturated sources per frame in the 3.6$\mu$m band decreasing to an average of one saturated source per frame in the 8.0$\mu$m band. 

{\it Timetable and Survey Tracking:} For our timetable, we give the
dates relative to the launch date ({\bf L+n} months).  The OSV data,
to validate the survey strategy for \glimpse, will be acquired at
approximately {\bf L+4}. Data acquisition for the full survey will
begin after OSV data are analyzed and the survey strategy is validated,  and will continue for about 9 months. The
observability of a section of the Galactic plane as a function of
Galactic longitude and time of year is shown in Figure 4.

%\begin{figure}
%\figurenum{4}
%\plotone{fg4_paper1.eps}
\begin{figure}[ht!]
\begin{center}
\includegraphics[angle=0,totalheight=\smpicsize]{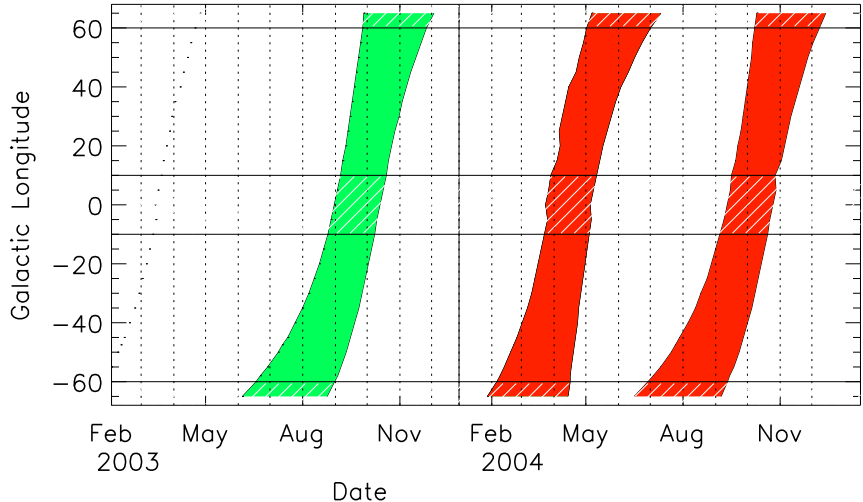}
\caption{\small{Visibility of the Galactic plane to \sirtf as a function of
date and Galactic longitude. This can be used to judge
when a particular part of the \glimpse region is likely to be observed.}}
\end{center}
\end{figure}

The first installment of the \glimpse Point Source Catalog will be
delivered to SSC at {\bf L+9}; the first installment of the mosaicked
data and \glimpse Point Source Archive will be delivered at {\bf
L+15}. Updates to each of these data products will be provided at six
month intervals after the first release dates.  The final version of
all \glimpse data products will be delivered to SSC at {\bf L+27}.
Documentation of the \glimpse survey will be available shortly after
launch ({\bf L+2}); updates will be provided with the data
releases. A preliminary version of the \glimpse Legacy Science Data
Products document is already available on our web site and will be available from the \sirtf Science Center.

\subsection{Data Processing}

The SSC will deliver Basic Calibrated Data (BCD) from the \glimpse IRAC observations to the \glimpse
team.  BCD will have gone through the following steps: validation,
addition of  header keywords, sense of InSb flux flipped, conversion to floating
point, correction for dichroic flip, normalization by Fowler number and
barrel-shifts, corrections of electronic bandwidth limitations, subtraction of dark/bias frames, correction for multiplexer bleed, correction for 
first-frame effects, linearization, flattening, detection of radiation hits,
subtraction of sky darks, flux calibration,  and detection of latents.  The positional
information will be good to $1.4\arcsec$ and the photometric accuracy
should be better than 10\% early in the mission. BCD will also contain
several ancillary data files, including the raw data, and several mask
images that contain information on cosmic rays, linearity corrections,
etc. A complete description of the processing that goes into
generating the BCD is given in \S 6.3 of the \sirtf Observers' Manual
(SIRTF 2002).

The \glimpse team will further process these data using an automated
pipeline (a ``Post-BCD'' pipeline) to correct for remaining
instrumental artifacts, extract and cross identify point sources, and mosaic the images. 
The resulting \glimpse Point Source Catalog and Archive and the set of mosaicked
images will  be released to the astronomical community via the SSC.  These released data will be more useful than the original BCD data, since they will benefit from the \glimpse team's experience in analyzing IRAC data in crowded and confused fields which have bright and positionally varying background emission. 

The data reduction will occur at the
University of Wisconsin-Madison using a network of LINUX workstations. The \glimpse pipeline is parallelized and will
simultaneously use multiple processors; data flow will be controlled
using OPUS pipeline software (Swade \& Rose 1999). Locally, the data
will be stored using the commercial database system, {\it Oracle 9i}, Release 2. 

We have divided the Post BCD pipeline into a few key levels with clearly defined steps within each. In order, these levels are:

\begin{itemize}

\item {\bf Data Verification: } Observed AORs are verified and checked to make sure the AOR was properly executed. The data are checked to verify that all SSC pipeline steps were carried out and checked for simple artifacts such as excessive cosmic ray hits and instrumental and down-link problems. A Quick-Look Validation Tool (QLVT) is used for spot-checks of frames and to inspect
frames with flagged problems.\footnote{ Quick-Look Validation Tool
(QLVT) is a quality assessment tool developed by team member Mark Wolfire, which
simultaneously displays four IRAC frames, data masks, and ancillary
2MASS/MSX data. The interface allows a user to insert comments and
mark potentially bad pixels.}
 
\item {\bf Basic Processing and Mask Propagation:} The data are corrected for the 
zodiacal background (Gorjian, Wright, \& Chary 2000). Pixels affected by stray light, 
banding\footnote{Banding refers to streaks that pre-launch tests
suggest might appear in the rows and columns radiating away from
bright sources in the 5.8 and 8.0 $\mu$m bands. The severity of this effect will be
determined on-orbit.}, or that are in the wings of saturated sources are flagged and corrected wherever possible. An error mask for these pixels is created. 

\item {\bf Point Source Extraction}:  Flux density and position of point sources are determined using DAOPHOT
(Stetson 1987). Positions of the brighter sources are checked against the 2MASS
Point Source Catalog.  Statistics  are computed for the residual images and used to assess the extraction process. Flux calibration is also spot-checked.

\item {\bf Bandmerging:} The point source lists obtained in the eight pointed observations (two passes each in four bands) are merged to produce the input for the generation of the \glimpse Point Source Catalog and \glimpse Point Source Archive.

\item {\bf  Mosaicked Image Production:} The data are resampled,
registered to a Galactic coordinate system,  and IRAC frames are background matched, if necessary.  The resulting images are turned into tiles of mosaicked images of $20\arcmin \times 20\arcmin$. The pixel size for these images is not yet finalized but will be approximately $0.6\arcsec$. 

\item {\bf Point-Source Photometry on the Mosaicked Image:} The mosaicked
images are used  to perform point-source photometry. The resulting source list is then bandmerged with source list from single-frame data.

\item{\bf \glimpse Point-Source Catalog and Archive Generation:} Sources found in mosaicked and single frames are cross identified. Appropriate quality and reliability filters are applied to generate the \glimpse point-source products. 

\end{itemize}

\subsection{Data Products}

There are four principal data products that will result from the
\glimpse program. These are:

\begin{enumerate}{}

\item A \glimpse Point Source Catalog (GPSC, or the ``Catalog''). The
flux limit for this catalog will be determined by the requirement that
the reliability be $\ge$99.5\%.  We currently estimate this flux
limit to be 1.1 mJy at $3.6 {\rm \mu m}$ and 2.5 mJy at $8.0
{\rm \mu m}$. The $8 {\rm \mu m}$ channel has a brighter limit due to the
increased diffuse background from PAH emission near 7.7 $\mu$m in the Galactic plane.
The Catalog photometric uncertainty will be $<$ 0.2 mag. For each IRAC band,
the Catalog will provide fluxes (with errors), positions (with
errors), the density of local point sources, the local sky
brightness, and flags that provide information on source quality and
any anomalies present in the data.  The Catalog is expected to contain
$\sim10^7$ objects.

\item A \glimpse Point Source Archive (GPSA or the ``Archive''),
consisting of point sources with signal levels $ \ge 5\sigma$
above the local background, to approximately a flux limit of 0.2-0.4 mJy. The photometric uncertainty is expected to
be $<$ 0.2 mag. The information provided will be the same as for the
Catalog. The Archive will contain $\sim5 \times 10^7$ objects.

\item Mosaicked Images for each band, each of approximately
20\arcmin$\times$20\arcmin\ angular coverage.  About 9000 FITS
formatted images will be tiled to smoothly cover the entire survey
area, using a Galactic coordinate system. The pixel resolution has not been finalized but will be about 0.6\arcsec.

\item The Web Infrared Tool Shed (WITS), a web interface to a
collection of model infrared spectra of dusty envelopes and
photodissociation regions (PDRs), updated for IRAC and MIPS band
passes. WITS currently resides on servers at the Infrared Processing
and Analysis Center (IPAC, www.ipac.caltech.edu).  The interface
contains two ``toolboxes'': DIRT (Dust InfraRed Toolbox) and PDRT
(PhotoDissociation Region Toolbox) which provide databases of
circumstellar shell emission models and PDR emission models. Users can
input data and retrieve best fit models. DIRT output includes central
source and dust shell parameters; PDRT output consists of gas density,
temperature, incident UV field and IR line intensities.

\end{enumerate}

\section{\glimpse Scientific Challenges}

The \glimpse project will produce a rich dataset that can be used for numerous and diverse investigations. Here we discuss some of the expected scientific uses of this survey and several complementary datasets. We first discuss the principal science goals of the \glimpse team: a census of star formation in the inner Galaxy and a study of Galactic structure as determined by the distribution of stars. This is followed by a discussion of community science and complementary data sets. 

\subsection{\glimpse Team Science Goals}

The \glimpse survey will uncover for the first time a huge number of stars in the inner Galaxy. As a result, it will be the survey of choice for those interested in the stellar structure of the Galaxy. Moreover, since the preponderance of star formation in the Galaxy is expected to occur in the inner Galaxy, studies using \glimpse will be able to characterize star formation in a wide range of environments. These are the two principal goals of the \glimpse science team. 

\subsubsection{Star Formation in the Inner Galaxy}

The \glimpse team will address several fundamental questions regarding star formation in the inner Galaxy using the \glimpse data products. These include the following: 

{\it At what rate are stars  forming in the inner Galaxy?}  Conservatively, we expect that analysis of  \glimpse data will reveal several thousand star formation regions (SFRs).  A search of the SIMBAD database lists over one thousand HII regions in our survey area; MSX 8${\rm \mu m}$
images of the \glimpse survey area indicate that there are many more
 to be found (Cohen \& Green 2001). 
Extrapolation from the luminosity functions of OB associations in other galaxies (McKee \& Williams 1997) suggest that there should be several thousand SFRs in our survey
area.  SFRs and other Galactic clusters are about 0.7 pc in diameter (Harris \& Harris 2000) and will
subtend $\ge10\arcsec$ at 15 kpc. Nearby clusters will
spread across arcminutes, and will be partially resolved
into individual stars and protostars. CO, HI, and radio continuum surveys (see \S 3.4) will also allow us to carry out targeted searches for star formation in the inner Galaxy.

{\it What are the spatial and mass distributions of lower mass stars in massive star formation regions?}  Because of high extinction and the low luminosity of low-mass stars in massive star formation regions, we know very little about the spatial and mass distributions of low mass ($M < 1-2 M_{\sun}$) stars associated with massive star formation regions. \glimpse data can be used to delineate both the spatial and mass distribution of lower mass stars in nearby (few kiloparsecs) star formation regions. 

{\it How does star formation vary as a function of position in the Galaxy?} Since \glimpse will allow for an unbiased sample of massive star formation in the inner Galaxy, it will allow us to search for variations in star formation properties, i.e., cluster density, initial mass function, gas content, in a wide range of Galactic environments. 

{\it How many low mass star formation regions in the inner Galaxy have been hidden until now?}  SFRs that contain only stars later than spectral type B3 are not easily detected in radio
continuum searches but will be detected by \glimpse. In the nearest SFRs, \glimpse data could be used for a census of the properties of intermediate mass pre-main sequence stars. 

{\it How does the infrared emission of star formation regions change over time?}  \glimpse data will provide information on 
the stellar content of all of the principal stages of massive star
formation, summarized recently in Churchwell (2002). Figure 5 demonstrates that these stages are expected to have significantly different IRAC colors. 

%\begin{figure}
%\figurenum{5}
%\plotone{fg5_paper1.eps}
\begin{figure}[ht!]
\begin{center}
\includegraphics[angle=0,totalheight=\tinypicsize]{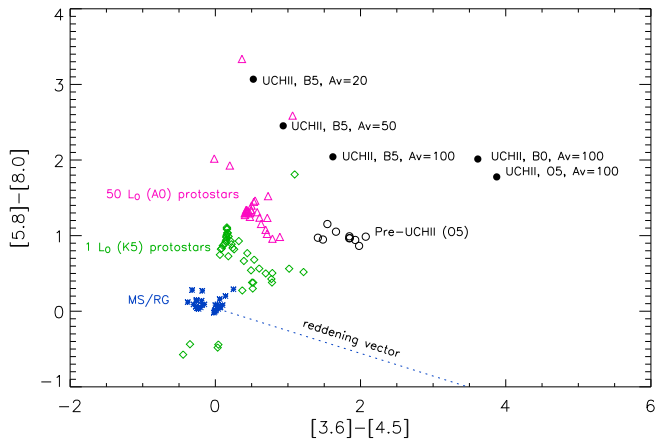}
\caption{ \small{IRAC color-color diagrams of model protostars 
calculated by Whitney et al. (2003) with both axes in magnitudes. Filled circles are UCHII models,
with various values of $A(V)$. The open circles are for a model of a
UCHII precursor (O star embedded in an infalling envelope and
accretion disk). Several viewing inclinations are plotted for this
model. Open triangles show different inclination models for
intermediate mass protostars (here, A0) from early protostar stage to
remnant disk. Open diamonds are similar models for low mass pre-main
sequence stars (K5). Only sources that are detectable to a distance
of 2 kpc are plotted (i.e., the more pole-on, less reddened
sources). The main sequence and giant branch are plotted as asterisks
(MS), with the reddening line shown by the dotted line. Reddening is
actually ``blueing'' in the longer wavelength bands, based on the ISM model of Li \& Draine (2001) because of the strong
silicate absorption features usually observed in these objects.  \glimpse will improve our understanding of the opacity in this wavelength range. }}
\end{center}
\end{figure}

\subsubsection{Galactic Structure}

There are three main questions that the \glimpse team seeks to address using \glimpse data: 

{\it Does the Galaxy have a stellar ring?}  The Galactic molecular ring contains
some 70\% of all molecular gas in the Galaxy, and should be the
dominant star-forming structure in the Milky Way (Clemens  et al. 2001).  On
this basis, Kennicutt (2001) suggests that our Galaxy should be
classified as an SB(r)bc pec.  Where in the ring are the stars forming? 
How do the properties of the gas correlate with star formation? 
How does the ring's star formation efficiency compare with starburst
regions in other galaxies? Comparison of \glimpse data with the
$^{13}$CO maps of this region of the Galaxy should yield answers to
these important questions.

{\it What are the nature of the spiral arms and disk in the inner
Galaxy?}  Observations have given clues for the gas (HI and CO), but
we know little about the stars and SFRs. Are stars formed on the
leading or trailing edges of gas arms?  How do the stars formed in arms
and in interarm regions differ? Drimmell \& Spergel (2001) show that K
band (stellar light) profiles are consistent with a two-armed
logarithmic spiral model, while the 240${\rm \mu m}$ (dust emission)
is consistent with a four-armed HII region distribution (Taylor \& Cordes 1993). \glimpse data will allow us to determine the
positions of individual star formation regions, account for regions of
high obscuration, and determine if the integrated light observed by
COBE/DIRBE is dominated by individual objects. \glimpse data will also allow us to identify different tracer populations (SFRs, OH/IR stars, IR carbon stars, etc) and their spatial distributions. The resulting information will help test models of gas dynamics, star
formation and evolution in the inner Galaxy (Englmaier \& Gerhard
1999). 

\glimpse data will also help  constrain values of the
 scale-lengths for the thin and thick stellar disks. Measuring the
ratio of the thin-to-thick disk scale-length will constrain the merger
history of the Galaxy (Quinn, Hernquist \& Fullager 1993).
Measuring the scale-length of the thin disk will establish whether the central mass distribution of the Galaxy is
stellar- or dark matter-dominated.  

{\it What are the principal properties of the central stellar bar of the Galaxy?}  COBE/DIRBE
data have shown the global distribution of the bar (Freudenreich 1998; Gerhard 2002). 
2MASS studies using IR carbon stars have also traced the structure and possibly the age of the Galactic bar (Cole \& Weinberg 2002).  \glimpse data will allow us to extend these studies, look for star formation at the ends of the bar, and explore the connection between the bar and inner spiral arms. 

\subsection{Community Science from \glimpse Data}

The wide variety of objects contained in the \glimpse survey region is illustrated in
Figure 6, which shows the positions of many types of objects overlaid on an
$8 {\rm \mu m}$ MSX map of a section of the \glimpse survey region. With the higher
sensitivity and angular resolution of \glimpse, together with the
color-select possibilities of seven or more photometric bands (\glimpse +
2MASS+MSX), data from \glimpse will allow the astronomical community to
evaluate the statistics, spatial distribution, and internal structures of numerous classes
of Galactic objects as well as providing new probes of the interstellar
medium. These include studies of stellar populations, photo-dissociation regions (PDRs), extinction, as well as serendipitous discoveries. 

%\begin{figure}
%\figurenum{6}
%\plotone{fg6_paper1_lowres.eps}
\begin{figure}[ht!]
\begin{center}
\includegraphics[angle=0,totalheight=\tinypicsize]{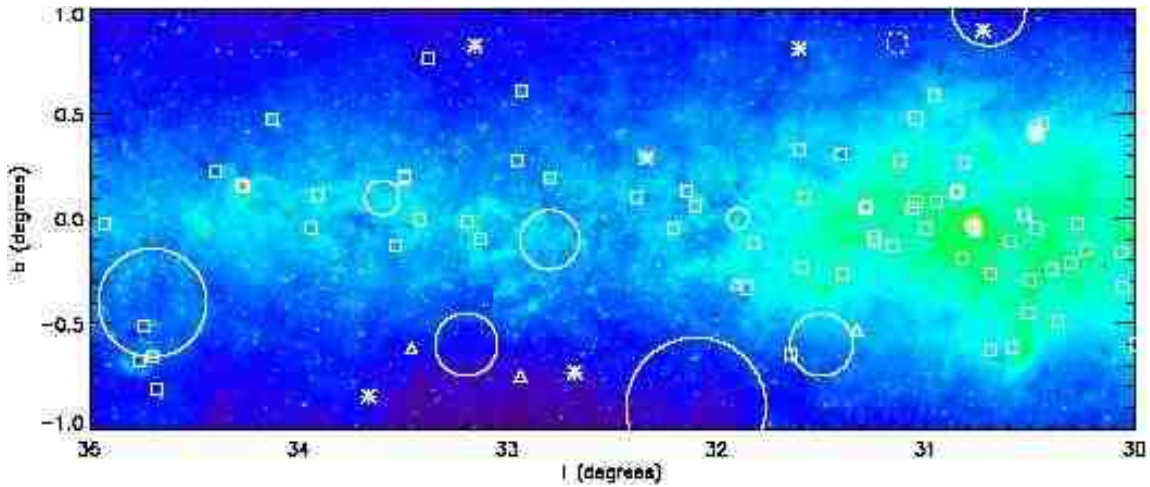}  
\caption{\small{MSX 8${\rm \mu m}$ image of a ten square degree piece of the \glimpse survey
region, with the positions of the following classes of objects
noted: OB stars (asterisks), HII regions (squares), planetary nebulae
(triangles), supernova remnants (solid circles), and open clusters
(dotted circles). Most of the diffuse emission in this image is
due to polycyclic aromatic hydrocarbons (PAHs). The prominent
star formation region at ($l=30^{\circ}.8$, $b=0.^{\circ}0$) is W43,
which is obscured optically ($A_{V}\approx 30$ mag). The
angular resolution of MSX was $18 \arcsec$ as compared to the
$1.2\arcsec$ pixel resolution expected from \sirtf/IRAC.}}
\end{center}
\label{fig6}
\end{figure}

{\it Stellar population studies}: Using a combination of 2MASS and
MSX data towards a sample of previously classified objects in the
Large Magellanic Clouds, Egan et al. (2001) showed that the
mid-IR $8 {\rm \mu m}$ band provides an important
``lever arm'' that allows color separation of many classes of
objects. Planetary nebulae, HII regions, and some
classes of C- and O-rich AGB stars have very red mid-IR colors.  Other
attempts to develop mid-IR and near-IR color selections focus on
infrared carbon stars using $(J-K_{s})$ vs. $K_{s}$ (Cole \& Weinberg 2002),
young stellar objects (YSOs) using ISOGAL [7]-[15] vs. [15] (Felli et
al 2002), brown dwarfs (Burrows et al. 1997), and carbon stars, OH/IR
stars, PN, Herbig AeBe stars, compact HII regions, and massive YSOs
using a combination of 2MASS J, H, K, and MSX $8 {\rm \mu m}$ (Lumsden et al
2002).

{\it Photo-Dissociation Regions (PDRs):} The near/mid IR spectrum of
photo-dissociation regions at the surface of molecular clouds is
dominated by emission bands at 3.3, 6.2, 7.7, 8.6, and 11.3 ${\rm \mu m}$,
probably arising from polycyclic aromatic hydrocarbon (PAH)
molecules (Peeters et al. 2002 and references therein). Figure 6 shows that  this emission is a
striking characteristic of the MSX maps of the Galactic
plane. \glimpse data can be used to characterize the spatial distribution of
different charge states of PAH to constrain the chemistry and
evolution of PDRs. The IRAC 3.6 ${\rm \mu m}$ band is sensitive to the
3.3${\rm \mu m}$ feature from neutral PAHs; the 5.8 and 8.0 ${\rm \mu
m}$ bands are sensitive to PAH$^{+}$, while the 4.5${\rm \mu m}$ band
contains no PAH features and thereby monitors the continuum (Bakes et al. 2001).

{\it Turbulence and Structure in Star Formation Regions:} A comparison of the infrared brightness fluctuations in star formation regions with the spectral line information from CO and HI observations can yield information about the source of ISM turbulence, by using the Velocity Channel Analysis technique of Lazarian \& Pogosyan (2000).  

{\it Interstellar Extinction:} Data from \glimpse will allow studies
of interstellar reddening in dense dusty regions and diffuse
environments, using the colors of stars that lie behind dark clouds.
This will allow testing near/mid infrared extinction models, two of which are given in Table 3
(Li \&
Draine 2001; Lutz et al. 1996). Since extensive grain coagulation
occurs in the inner regions of dense clouds, the IR extinction law
could be quite different between very dense clouds like the MSX dark clouds (Egan et al. 1998) and presently observable
regions. \glimpse data will allow a vital characterization of the
variation of extinction properties with environment.

{\it Serendipity:}  Since the inner Galaxy is the region of the sky with the greatest extinction, it is also the direction in which one is most likely to make serendipitous discoveries. The recent 2MASS discoveries of new globular clusters in the Galactic plane (Hurt et al. 2000) and galaxies in the ``Zone of Avoidance'' (Jarrett et al. 2000)  hint at the possibilities for \glimpse.

\subsection{Complementary Data Sets}

The value of the \glimpse data products will be enhanced by the
availability of complementary IR and radio data sets. The basic
characteristics of the IR surveys are given in Table 4. Data sets that
we anticipate will be the most useful, and that will play important
roles in the \glimpse team science studies, are

\begin{enumerate}{}

\item {\it 2MASS}: This survey (Cutri et al. 2001) provides an ideal
companion dataset to \glimpse, with an excellent match in both
sensitivity and angular resolution for many types of objects. Many
objects will have \glimpse$+${\it 2MASS} magnitudes in a total of seven near-IR and mid-IR
bands!  This will allow for a wide variety of different possible color
selections and SED's from $\sim$ 1 $\mu$m to 8 $\mu$m. The \glimpse
bands provide crucial mid infrared information.

\item{ \it MSX}: The MSX (Price et al. 2001) survey provides a good
match to the \glimpse/IRAC 8$\mu$m band. The MSX dataset will be
particularly useful for studies of diffuse emission. It provides a good
complement to the \glimpse data for bright sources, since the
saturation limit of \glimpse is only slightly brighter than the faint
detection limit for MSX.

\item {\it Arecibo/Green Bank Telescope/Australian Telescope Compact Array Surveys of \glimpse
HII Regions:} This dataset resolves the 
distance ambiguities to many massive star formation regions. The data
include $>100$ objects with resolved distance ambiguities which will be 
published and made available on the \glimpse web
site: www.astro.wisc.edu/glimpse.  

\item {\it Milky Way Galactic Ring Survey (GRS):} A Boston University and
Five College Radio Astronomy Observatory collaboration, this is a
large-scale ${\rm ^{13}CO~J=1 \rightarrow 0}$ molecular line survey of the inner Galaxy between latitudes $-1$\deg\ to $+1$\deg\ and 
longitudes 18\deg\ and 52\deg, with an angular resolution of $22
\arcsec$ and a velocity resolution of $0.3~ {\rm km~s^{-1}}$ (Simon et al. 2001) It is available
through the GRS website: www.bu.edu/grs. GRS will be completed by winter 2003. 

\item {\it The International Galactic Plane Survey:} This survey will map
the Milky Way disk in the HI 21-cm line with a resolution of 1\arcmin\
and ${\rm 1~km~s^{-1}}$ over the entire \glimpse survey area.  The data cubes
will be available at www.ras.ucalgary.ca/IGPS. The Southern Galactic Plane Survey is available at ftp://ftp.astro.umn.edu/pub/users/john/sgps .

\end{enumerate}

There are several other surveys that will provide a useful complement
to \glimpse.  These include the ISOGAL survey at 7 and 15 $\mu$m
(Omont et al 2003; Felli et al 2002) which provides complentary data
for the inner Galaxy region not covered by the \glimpse survey, and
the planned all-sky ASTRO-F survey at 8.5 -- 175 $\mu$m which will provide
provide an extension to large galactic latitudes and 
longitudes, although at lower resolution and sensitivity than
\glimpse (See Table 4).  High angular resolution X-ray surveys of the 
Galactic plane using Chandra (Grindlay et al 2003; 
http://hea-www.harvard.edu/ChaMPlane) and XMM (Helfand et al 2002) will also provide a useful comparison to \glimpse data in selected regions of the Galactic plane. 

\section{Summary}

The \glimpse project will allow us to study, for the first time, the
stellar content of the inner Galaxy with high angular resolution and a
minimum of extinction. The \glimpse team and others will use these
data to study Galactic stellar structure, characterizing the stellar
content and star formation in the Galactic bar and inner spiral
arms. It may allow us to ascertain whether the Galaxy is a ringed
spiral.  In addition, the data will be used to study the distribution
and statistics of star formation throughout the Galaxy.

The survey will use the IRAC instrument on SIRTF to image 220 square degrees in four bands (3.6,4.5,5.8, and 8.0 $\mu$m) with a pixel resolution of 1.2\arcsec .   It will cover two strips spanned by $|b|\le 1^{\circ}$ and $|l|=10^{\circ}$ to $65^{\circ}$, a region covering the outer ends of the Galactic bar, the Molecular ring, and four spiral arm tangencies. The resulting dataset will be  the most panoramic  produced by SIRTF. 

The principal data products from the \glimpse team will be a high-reliability \glimpse Point Source Catalog (GPSC) with about ten million sources and approximate flux limit of 1.0 mJy (3.6$\mu$m band) to 2.5 mJy (8.0 $\mu$m band), a  \glimpse Point Source Archive (GPSA) with about 50 million sources and an approximate flux limit of 0.2-0.4 mJy , a set of mosaicked images for each band, and a set of Web based analysis tools. The first release of the GPSC will be nine months after the launch of \sirtf, and the first installments of the GPSA and mosaicked images will be be fifteen months after launch. These data products and the supporting documentation will be updated at six month intervals and will be complete 27 months after launch. 

\glimpse data will drive  a wide range of scientific investigations including the search for rare, bright Galactic objects, stellar population studies, studies of Galactic structure, high angular resolution studies of diffuse emission in PDRs, and studies of extinction in the near to mid-infrared. The science from \glimpse data will fuel observing programs and scientific
investigations for decades. The probability of serendipitous discoveries is high for the \glimpse
survey. We expect that it will lead to the discovery of new stellar clusters and galaxies hidden behind what had previously been an impenetrable wall of dust. We eagerly look forward to providing this resource to the
community.

\acknowledgments Support for this work, part of the Space Infrared Telescope Facility (SIRTF) Legacy Science Program, was provided by NASA through an award issued by the Jet Propulsion Laboratory, California Institute of Technology under NASA contract 1407.  This research has made use of the SIMBAD database, operated at CDS, Strasbourg, France, the NASA/IPAC Infrared Science Archive, which is operated by the Jet Propulsion Laboratory, California Institute of Technology, under contract with NASA, and data products from the Midcourse Space Experiment. Processing of the Midcourse Space Experiment data was funded by the Ballistic Missile Defense Organization with additional support from NASA Office of Space Science. 

\input{tab1_paper1.tex}
%\clearpage
\input{tab2_paper1.tex}
%\clearpage
\input{tab3_paper1.tex}
%\clearpage
\input{tab4_paper1.tex}

%% The following command ends your manuscript. LaTeX will ignore any text

%% The following command ends your manuscript. LaTeX will ignore any text
%% that appears after it.

\end{document}

%% file: tab1_paper1.tex
\begin{deluxetable}{lcl}
\tabletypesize{\scriptsize}
\tablecaption{Objects in the GLIMPSE survey area\tablenotemark{a}}
\tablehead{ \colhead{Object Type} & \colhead{Number known in GLIMPSE region}  & \colhead{Examples} }
\startdata
MSX point sources       &  61,321      &   \\
IRAS point sources      &  15,501      &  \\
HII regions             &  1174        & M16 (Eagle Nebula), M17, W43,W49,W51 \\
ROSAT point sources     & 459          &                                    \\
Radio pulsars           & 264          &                                    \\
Dark clouds             & 210          & Coalsack, Vulpecula Rift, B48, LDN485 \\
Galaxies                & 157          & IRAS 16232-4917, GAL 312.11-0.20   \\
ASCA point sources      & 144          &                                    \\
Supernova remnants      & 100          & Kes 69, RCW 103, CTB 37A/B, Carina \\
O/B stars               & 98           & BD-20 5020, BD-15 4930,BD+31 3921  \\
Open clusters           & 76           & NGC 3572, Sco OB2, Sct OB2, Westerlund 1   \\
Planetary nebulae       & 65           & NGC 6537, NGC 6842, IC 4637        \\
Wolf-Rayet stars        & 50           & IC14-17, Vyl-3, W43\#1, The 3      \\
Globular clusters       & 1            & 2MASS-GC01                        
\enddata
\tablenotetext{a}{A listing of the objects in the \glimpse region with links to various databases and maps showing the positions of these objects can be found at the \glimpse website: http://www.astro.wisc.edu/glimpse .} 
\end{deluxetable}

%% file: tab2_paper1.tex
\begin{deluxetable}{ll}
\tabletypesize{\scriptsize}
\tablecaption{Summary of Principal GLIMPSE Characteristics}
\tablehead{ \colhead{Characteristic} & \colhead{Description}}
\startdata
Galactic Longitude Limits            & $|l|=10^{\circ}-65^{\circ}$   \\
Galactic Latitude Limits             & $|b|<1^{\circ}$   \\
Total Survey Area                    & 220 square degrees   \\
Total Survey Time                    & 400 hours \\
Total Resolution elements per band   & $\sim 2 \times 10^{9}$   \\
Total Number of IRAC frames per band & $\sim$80,000 \\
IRAC frame size                      & $5.17\arcmin \times 5.17\arcmin$ ($256 \times 256$ pixels) \\
Pixel resolution               & $1.2\arcsec \times 1.2\arcsec$ \\
Frame time per visit                 &  2 seconds \\
Number of visits per position        &  2 \\
Frame overlap\tablenotemark{a}                        & $14.4\arcsec$ (12 pixels) \\
IRAC wavebands                       &  3.6$\mu$m, 4.5$\mu$m, 5.8$\mu$m, 8.0$\mu$m \\
$5\sigma$ Sensitivity (4 sec.) & 0.2, 0.2, 0.4, 0.4 mJy \\
Estimated completeness limit for GPSC\tablenotemark{b} & 1.0, 1.5, 2.0, 2.5 mJy \\
Saturation limits              & 180, 190, 570, 470 mJy \\
Galactic features covered      & Outer ends of Galactic bar, molecular ring, \\
                               & four spiral arm tangencies: Norma($l=333^{\circ}$), \\
                               & Scutum-Crux($l=30,320^{\circ}$), Sagittarius-Carina($l=50^{\circ}$)
\enddata
\tablenotetext{a}{Subject to change after observing strategy validation period.}
\tablenotetext{b}{Flux level necessary to achieve 99.5\% reliability based on simulated data. These values are subject to change after the observing strategy validation period.}
\end{deluxetable}

%% file: tab3_paper1.tex
\begin{deluxetable}{lrrrrcrcrr}
\tabletypesize{\scriptsize}
\tablecaption{GLIMPSE/IRAC, 2MASS, and MSX Characteristics}
\tablehead{ \colhead{Survey/Band} & \colhead{$\lambda$}  & \colhead{Bandwidth}& \colhead{Zero mag.\tablenotemark{a}} 
& \colhead{5$\sigma$ Sens.}  & \colhead{Complete Lim.} & \colhead{Sat. Lim} &   \colhead{$A_{\lambda}/N(H)$\tablenotemark{e}} 
& \colhead{$A_{\lambda}/A_{V}$\tablenotemark{e}} &  \colhead{$A_{\lambda}/A_{V}$\tablenotemark{f}} \\ 
\colhead{} & \colhead{($\mu$m)}& \colhead{($\mu$m)} & \colhead{(Jy)} & \colhead{(mJy)} & \colhead{(mJy)} &
\colhead{(mJy)} & \colhead{($10^{-22}~{\rm cm^{2}~mag}$)} & \colhead{} & \colhead{}
 }
\startdata
2MASS/J      & 1.24 & 0.25 & 1592  &  0.4      & 0.3--0.8   & 15920 & 1.482 & 0.293 & 0.293 \\ 
2MASS/H      & 1.66 & 0.30 &1024  &  0.5      & 0.4--0.9   & 10240 & 0.959 & 0.190 & 0.190 \\ 
2MASS/K      & 2.16 & 0.32 & 667  &  0.6      & 0.5--1.3   &  6670 & 0.593 & 0.117 & 0.117 \\ 
%IRAC 1 & 3.548 &  278  &  0.2      & 0.4--1.0   &   180 & 0.237 & 0.047 & 0.058 \\ 
%IRAC 2 & 4.492 &  180  &  0.2      & 0.7--1.5   &   190 & 0.153 & 0.030 & 0.053 \\ 
%IRAC 3 & 5.656 &  117  &  0.4      & 1.0--2.0   &   570 & 0.103 & 0.020 & 0.053 \\ 
%IRAC 4 & 7.835 &   63  &  0.4      & 1.2-2.5    &   470 & 0.161 & 0.032 & 0.053 \\
GLIMPSE/IRAC 1 & 3.55 & 0.66 & 289  &  0.2 \tablenotemark{b}     & 0.4--1.0\tablenotemark{c}          &   180\tablenotemark{d}  & 0.237 & 0.047 & 0.058 \\ 
GLIMPSE/IRAC 2 & 4.49 & 0.88 & 183  &  0.2 \tablenotemark{b}     &  0.7--1.5\tablenotemark{c}         &   190\tablenotemark{d} & 0.153 & 0.030 & 0.053 \\ 
GLIMPSE/IRAC 3 & 5.66 &  1.32 & 131  &  0.4 \tablenotemark{b}     &  1.0--2.0\tablenotemark{c}         &   570\tablenotemark{d} & 0.103 & 0.020 & 0.053 \\ 
GLIMPSE/IRAC 4 & 7.84 &   2.40 & 71  &  0.4  \tablenotemark{b}    & 1.2--2.5\tablenotemark{c}          &   470\tablenotemark{d} & 0.161 & 0.032 & 0.053 \\ 
MSX/Band A  & 8.28 &   3.36 & 58  & 30 & 60 &   400000  & 0.022    & 0.044    & 0.053  \\ 
% Magnitude version of the above table...
%J     & 1.235 & 1592  &  16.5     & 16.8--15.8 & 5.0 & 
%H     & 1.662 & 1024  &  15.8     & 16.1--15.1 & 5.0 &  
%K     & 2.159 &  667  &  15.1     & 15.3--14.3 & 5.0 & 
%IRAC 1& 3.548 &  278  &  14.7     & 14.5--13.5 & 8.0 & 
%IRAC 2& 4.492 &  180  &  14.1     &   --       & 7.4 & 
%IRAC 3& 5.656 &  117  &  13.1     &  --        & 5.8 & 
%IRAC 4& 7.835 &   63  &  12.5     & 12--11     & 5.3 & 
%MSX A & 8.276 &   58  & $\sim8.2$ & $\sim 7.5$ &  -- & 

\enddata
\tablenotetext{a}{Zero magnitude for J,H,K from 2MASS Explanatory supplement. Zero magnitude from MSX from Cohen, Hammersley, \& Egan (2000). IRAC zero magnitudes are interpolated from Table 7.5 in Tokunaga (2000); these may differ by as much as 13\% from the final adopted values (M. Cohen, priv communication).} 
\tablenotetext{b}{5$\sigma$ sensitivity for a 4 second integration using IRAC.}
\tablenotetext{c}{Flux limits for the Point Source Catalog/Archive are based on assuming a reliability of $\ge 0.995$ for the Point Source Catalog and is based on simulations. Limits may change after observing strategy validation period.}  
\tablenotetext{d}{Diffuse source saturation limit for GLIMPSE/IRAC bands in units of MJy/sr can be obtained by multiplying the point source number (in mJy) by $30/\sqrt{N_{pix}}$, where $N_{pix}$ is the number of pixels in the point source PSF. $30/\sqrt{N_{pix}}=9.2$ for Band 1 and decreases to 7.0 for Band 4.}
\tablenotetext{e}{Extinction curve from Li \& Draine (2001)}
\tablenotetext{f}{Extinction curve from Lutz et al (1996)}

\end{deluxetable}

%% file: tab4_paper1.tex
\begin{deluxetable}{llllll}
\tabletypesize{\scriptsize}
\tablecaption{Summary of Infrared Surveys}
\tablehead{ \colhead{Survey} & \colhead{Wavebands}  & \colhead{Resolution} & \colhead{Coverage}  & \colhead{Sensitivity} & \colhead{Website}  \\
   \colhead{} &               \colhead{($\mu m$)}  & \colhead{($\arcsec$)} & & \colhead{} & }
\startdata
GLIMPSE  & 3.6,4.5,5.8,8.0           & $\le 2$           & $|l|=10$-$65^{\circ}$,$|b| \le 1^{\circ}$ &  0.2,0.2,0.4,0.4 mJy\tablenotemark{a} & 
  www.astro.wisc.edu/glimpse\\ 
2MASS     &1.22,1.65,2.16             & 2        & all-sky  & 0.4,0.5,0.6 mJy & 
   www.ipac.caltech.edu/2mass \\
DENIS      & 0.97,1.22,2.16          & 1-3 &$\delta=+2$ to $-88^{\circ}$ & 0.2,0.8,2.8 mJy &  
   cdsweb.u-strasbg.fr/denis.html \\
MSX	         & 4.1,8.3,12,14,21        & 18.3                &$ l=0$-$360^{\circ}$,$|b| \le 5^{\circ}$ & 10000,100,1100,900,200  mJy  & 
  www.ipac.caltech.edu/ipac/msx \\
ISOGAL	&7,15                              & 6                      &$|l| \le 60^{\circ}$,$|b| \le 1^{\circ}$\tablenotemark{b} & 15,10 mJy & 
www-isogal.iap.fr/  \\ 
IRAS          &12,24,60,100              & 25--100           &all-sky & 350,650,850,3000 mJy & 
 irsa.ipac.caltech.edu/IRASdocs \\ 
ASTRO-F\tablenotemark{c} & 8.5,20,62.5,80,155,175      & 5--44  & all-sky  & 20-100 mJy & www.ir.isas.ac.jp \\ 
COBE/DIRBE\tablenotemark{d} & 1.25--240 & $0.7^{\circ}$ & all-sky  & 0.01-1.0 ${\rm MJy~sr^{-1}}$ &  space.gsfc.nasa.gov/astro/cobe/ 
\enddata
\tablenotetext{a}{Best effort $5\sigma$ limit for sources in the GLIMPSE Point-Source Archive; the GLIMPSE Point-Source Catalog will contain only sources at about the 20$\sigma$ level to insure high reliability.}
\tablenotetext{b}{Survey contained only selected fields in this region, totaling 16 square degrees.}
\tablenotetext{c}{Launch planned February 2004}
\tablenotetext{d}{DIRBE photometric bands are 1.25, 2.2, 3.5, 4.9, 12, 25, 60, 100, 140, and 240 $\mu$m. We report the diffuse flux sensitivity rather than point source sensitivity due to the large beam size.}
\end{deluxetable}